\documentclass[onecolumn,showpacs,preprintnumbers,amsmath,amssymb,prl,preprint]{revtex4-1}
\usepackage{graphicx}
\usepackage{multirow}
\usepackage{bm}
\usepackage[breaklinks=true,colorlinks=true,linkcolor=blue,urlcolor=blue,citecolor=blue]{hyperref}
\usepackage{booktabs}

\graphicspath{{./figures/}}

\begin{document}
\title{Valley filtering effect of phonons in graphene with a grain boundary}

\author{Xiaobin \surname{Chen}$^{1,2}$}
\email{Email:chenxiaobin@hit.edu.cn}
\author{Yong \surname{Xu}$^3$}
\author{Jian \surname{Wang}$^{2,4}$}
\email{Email:jianwang@hku.hk}
\author{Hong \surname{Guo}$^{5,6}$}
\email{Email:guo@physics.mcgill.ca}

\affiliation{
$^1$State Key Laboratory on Tunable laser Technology and Ministry of Industry and Information Technology Key Lab of Micro-Nano Optoelectronic Information System, School of Science, Harbin Institute of Technology, Shenzhen 518055, China\\
$^2$Department of Physics and the Center of Theoretical and Computational Physics, The University of Hong Kong, Hong Kong, China \\
$^3$Collaborative Innovation Center of Quantum Matter and State Key Laboratory of Low Dimensional Quantum Physics, Department of Physics, Tsinghua University, Beijing 100084, People¡¯s Republic of China\\
$^4$The University of Hong Kong Shenzhen Institute of Research and Innovation, Shenzhen 518053, China\\
$^5$College of Physics and Energy, Shenzhen University, Shenzhen 518060, China\\
$^6$Department of Physics, 3600 University, McGill University, Montreal, Quebec H3A 2T8, Canada
}

\date{\today}
\begin{abstract}
Due to their possibility to encode information and realize low-energy-consumption quantum devices, control and manipulation of the valley degree of freedom have been widely studied in electronic systems. In contrast, the phononic counterpart--valley phononics--has been largely unexplored, despite the importance in both fundamental science and practical applications. In this work, we demonstrate that the control of ``valleys" is also applicable for phonons in graphene by using a grain boundary. In particular, perfect valley filtering effect is observed at certain energy windows for flexural modes and found to be closely related to the anisotropy of phonon valley pockets. Moreover, valley filtering may be further improved using Fano-like resonance. Our findings reveal the possibility of valley phononics, paving the road towards purposeful phonon engineering and future valley phononics.
\end{abstract}


\maketitle

\section{Introduction}

Owing to potential realization of low-dissipation devices, valleytronics has received considerable research interest. For electrons, degenerate but inequivalent valley states around the Fermi level form another degree of freedom besides the spin degree of freedom, providing a feasible way to design information-storage or -processing devices with low energy consumption. As the key factor for encoding information, valley degeneracy of electrons can be controlled by strain,\cite{Gunawan_PRL_2006} magnetic fields,\cite{Behnia_NatPhys_2012} optical pumping with circularly polarized light,\cite{TonyHenz_NatNano2012} substrates,\cite{FengJi_PRB2015} $etc$. Also, valley-polarized current can be achieved by cyclic strain,\cite{Jiang_PRL_2013} second-order nonlinear response,\cite{YaoWang_PRL_2014} temperature gradient,\cite{cxb_prb_2015,Jauho_PRL2015,YuZhizhou_Carbon2016} and valley filters.\cite{Beenakker_NatPhys_2007,Santos2009,Gunlycke_PRL_2011,xuFuming_NJP2016,Zheng_2D2017}

For valleys of phonons, however, it remains largely unexplored. Particularly, phonons cannot be manipulated directly by electric fields or magnetic fields due to the lack of spin and charge degrees of freedom. To lift the degeneracy of phonon valleys, one needs to break the time-reversal symmetry and inversion symmetry, which is rather complicated.\cite{xuyongPRB2017} Therefore, obtaining phonon valley polarization via valley filtering is an attractive and promising option.

For geometry-related valley filtering effects of electrons,\cite{Beenakker_NatPhys_2007,Santos2009} the key ingredient is to make a single-valley region to filter out one valley of electrons. In these cases, no symmetry is explicitly required. For phonons, geometry-related valley filtering should be applicable similarly.

One particular type of valley filters, as learned from electrons, is grain boundaries (GBs), which are ubiquitous in chemical vapor deposition samples. GBs represent a class of interfaces exhibiting exotic electronic properties such as quasi-1D metallic states\cite{Batzill_nat2010,ZouXL_small2015,ZouXL_AccChem2015} and selective valley filtering for electrons.\cite{Yazyev_NatMater_2010,Gunlycke_PRL_2011} The influence of GBs on thermal conduction of phonons has been widely studied.\cite{JAP2004_Keblinski,Kimmer_PRB2007,GuoJing_APL_2012,Pop_APL_2013} However, valley filtering effect of phonons in GB systems lacks of investigations.

Motivated by recent development of both valleytronics and phononics,\cite{YaoWang2016,LiBaowen_PRL_2004,Yizhou_PRL2017} in this work, we explore the valley degree of freedom in an entirely different type of quasi-particles, $i.e.$, phonons. We show that phonon valley states can be filtered by using a grain boundary. In addition to valley-selective transport, Fano-like resonance appears in the transmission spectrum, enhancing the valley polarization of phonons further.

\begin{figure*}
      \centering
      \includegraphics[width=0.8\textwidth]{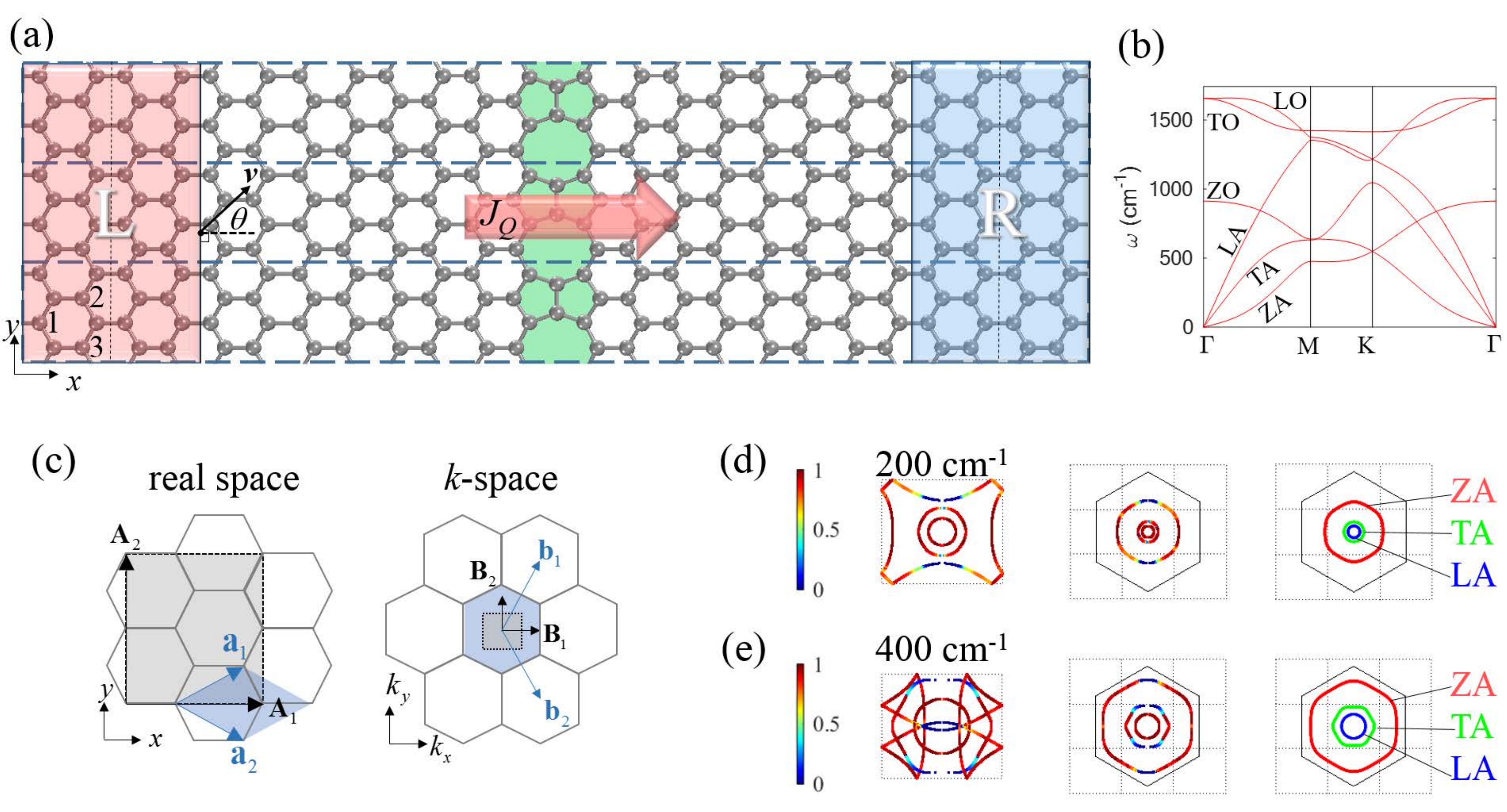}\\
      \caption{ (a) The atomic configuration used for phonon transport calculation. Three horizontal long blocks represent three unit cells along the $y$-direction. The red-shaded and blue-shaded areas on two sides mark the left (L) and right (R) thermal leads, which are semi-infinite crystalline graphene sheets. As divided by vertical dotted lines, each principal layer of both thermal leads consists of 8 atoms in a unit cell. (b) Phonon dispersion of graphene calculated using RESCU.\cite{Vincent2016} (c) ${\bf a}_{1,2}$ are primitive lattice vectors of graphene and ${\bf A}_{1,2}$ are lattice vectors of an 8-atom unit cell of graphene as shown in the L/R thermal leads in (a). ${\bf b}_{1,2}$ and ${\bf B}_{1,2}$ are the reciprocal lattice vectors of graphene and the 8-atom unit cell, with the first Brillouin zones (BZs) shaded in blue and grey, respectively.
      (d-e) $\bf k$-resolved transmission at $\omega=200$ and $400$ cm$^{-1}$ shown in the folded 1$^{\textrm{st}}$ BZ (left panel) and unfolded BZ (middle panel). In the middle panel, the folded BZs are also indicated in dotted boxes, covering the hexagonal unfolded BZ.
      Contour plots showing the branch information (ZA (red), TA (green), LA (blue), ZO (cyan), and LO (magenta)) are shown in the right panel.
      }\label{fig1}
      \end{figure*}

\section{Model and Methods}
We illustrate our idea by investigating the transport properties of phonons through a 558 GB in graphene. The 558 GB is an extended line defect [Fig. \ref{fig1}(a)] containing carbon-atom pentagons and octagons.\cite{Pop_APL_2013,Huaqing_PRB_2013} It is actually a 0$^\circ $ GB system,\cite{Mele_PRB_2015} where selective valley filtering of electrons has been shown previously.\cite{Yazyev_NatMater_2010,Gunlycke_PRL_2011} Graphene with a 558 GB was experimentally realized\cite{Batzill_nat2010} in 2010. Figure \ref{fig1}(a) shows the transport system, which consists of two semi-infinite thermal leads ($i.e.$, semi-infinite ideal graphene), and a central scattering region with a 558 GB. The structure is periodic along the transverse ($y$) direction.

   In this work, we focus on the quasi-ballistic quantum transport regime, neglecting phonon-phonon and electron-phonon interactions. In realistic experimental systems, the graphene layer might be encapsulated with hBN layers or be placed on some substrates. In this work we shall not consider extrinsic effects due to phonon coupling between the different materials, i.e., the interlayer coupling of phonons is ignored such that the graphene layer behaves as a free-standing one.

    \subsection{Mode-Matching Method}
     To investigate transport behaviors of individual phonon modes of graphene, we utilize the mode-matching (MM) method,\cite{Ando_PRB_1991,Kelly_prb_2005,Savito_PRB_2008}
    which has been widely used for investigating electronic transport\cite{lei,cxb_prb_2015} and was recently applied to phonon transport.\cite{ZhangGang_PRB_2015,Fisher_PRB2017}
    In this part, we give a brief introduction of the mode-matching method. (More details can be found in the Appendix A.)

   To apply the MM method, the transport model is artificially divided into``principal layers" along the transport direction and be labeled with layer number $\cdots -2, -1,0$,$1,2,\cdots,N$,$N+1, N+2,\cdots$ Here, $-\infty$ to $0$ label the left thermal lead, $1$ to $N$ label the central scattering region, and $N+1$ to $+\infty$ label the right thermal lead.
    Principal layers are usually composed of several unit cells to ensure that only nearest-neighbor interaction exists between them.
    For one mode $\psi_0$ going through the device from the left thermal lead, it becomes $\psi$ at the right thermal lead. This evolution can be expressed in terms of phonon Green's functions as:
    \begin{equation}
            \psi  = {\bf G}^r{\left( {{{\bf G}_0^r}} \right)^{ - 1}}{\psi _0}, \label{eq:psi=G*G0niPsi0}
            \end{equation}
    where ${{\bf G}^r}$ and ${{\bf G}_0^r}$ are the retarded Green's of the central region with/without coupling between the thermal leads and the central region.
    More precisely, for a right-propagating eigen mode ${{\bf{u}}_{Ln}}( + )$ of the left thermal lead, it evolves into
    \begin{align}
            {\psi _{l}} = {\bf G}_{n,0}^r{\left( {{\bf G}^{(0)r}_{L;0,0}} \right)^{ - 1}}{{\bf{u}}_{Ln}}( + )
    \end{align}
    at layer $l$. To find out the transmission function, one may look at the wave function at layer $N+1$, which is the left-surface of the right thermal lead. Decomposing the wave function into eigen-modes ${{\bf{u}}_{Rm}}\left(  +  \right)$ of the right thermal lead, we have
     \begin{align}
            {\psi _{N+ 1}} = \sum\limits_m^{} {{\tau _{nm}}{{\bf{u}}_{Rm}}\left(  +  \right)}.
            \end{align}
    And the transmission amplitude of mode $n$ from the left thermal lead to mode $m$ of the right thermal lead is
     \begin{align}
            t_{mn}
            & = \sqrt {\frac{{{v_{Rm}}{a_L}}}{{{v_{Ln}}{a_R}}}} {\tau _{mn}}.
            \end{align}
             where $v_{Rm(Ln)}$ is the group velocity of mode $m(n)$ in R(L), $a_{L/R}$ is the principal-layer lattice constant of the $L/R$ thermal leads.
   Consequently, transmission of mode $n$ (in L) to mode $m$ (in R) is
     \begin{align}
{\Xi_{mn}}=|t_{mn}|^2.
            \end{align}
   The total transmission function is obtained by adding up transmission coefficients of right-propagating modes:
        \begin{align}
        \Xi(\omega)=\sum_{m,n} \Xi_{mn}(\omega).
            \end{align}

    \subsection{Computational Details}
       Force constants of the transport system were obtained using the finite-displacement method\cite{Parlinski_PRL1997,AtsushiTogo_2010} as implemented in the first-principles Realspace Electronic Structure CalcUlator (RESCU).\cite{Vincent2016} Norm-conserving pseudo-potential for carbon atoms was used. Self-consistent calculations were performed within the local-density approximation (LDA) with energy criteria of $1\times10^{-6}$ Hartree and charge variation criteria of $1\times10^{-5}$.

       For the calculation of phonon dispersion of graphene, the optimized carbon-carbon bond length of 1.40787 \AA~was used. In addition, a 4$\times$4$\times$1 supercell structure and a real-space resolution of 0.4 Bohr were employed.

       For the calculation of the whole transport system, a $k$-mesh of 1$\times$3$\times$1 and a real-space resolution of 0.4 Bohr were chosen. The central region was relaxed until forces exerted on the central region were less than $1\times10^{-3}$ Hartree/Bohr. Then, a $1\times3\times1$ supercell of the transport system was constructed with periodic boundary condition for the calculation of force constants. The constructed system contains 294 atoms and requires calculations of forces for 392 displaced configurations.

       Transmission calculation based on the force constants was done using the PHonon MODEs (PHMODE) code. We adopted force constants of ideal graphene structure for the thermal leads to ensure that leads were in ideal crystalline structures. Then, for the utilization of the MM method, the transport system was divided into principal layers along the transport direction ($x$)\cite{Kelly_prb_2005} and a cutoff of 5~\AA~was applied to the force constants of the whole transport system to ensure that only nearest-neighbor interaction exists between principal layers.

    \subsection{Unfolding phonon band structures}\label{sec.unfold}

        By applying the mode-matching method, we can obtain mode-resolved transmission coefficients. However, the principal layers in leads are not primitive unit cells. What we solve from the mode-matching method are eigenvectors in the supercell and wave vectors $\bf K$ in the corresponding folded Brillouin Zone (BZ). To fully explore the mode dependence of transmission, it is vital for us to unfold the $\bf K$ to those $\bf k$-vectors within the $1^{\textrm{st}}$ BZ of the primitive cell.\cite{Huaqing_NJP_2014,ZhengFawei_CPC_2016} Here, we propose a rather convenient way to realize this idea.

        According to Bloch's theorem, it is well known that wave function of a periodic system $\psi_{n{\bf k}}({\bf r})$ satisfies the relation: $\psi_{n{\bf k}}({\bf r+R})=\psi_{n{\bf k}}({\bf r})\exp(i{\bf k}\cdot {\bf R})$ for all $\bf R$ in the Bravais lattice, where $\bf k$ is the wave vector.
        As shown in Fig. \ref{fig1}(a) of the main text, we can pick up three equivalent atoms, for example, those labeled with 1, 2, 3, in a principal layer of the lead L. Then, we have
        \[\begin{array}{l}
        {e^{i\left[ {{k_x}\left( {{x_2} - {x_1}} \right) + {k_y}\left( {{y_2} - {y_1}} \right)} \right]}} = {u_{n;2\alpha }}/{u_{n;1\alpha }},\\
        {e^{i\left[ {{k_x}\left( {{x_3} - {x_1}} \right) + {k_y}\left( {{y_3} - {y_1}} \right)} \right]}} = {u_{n;3\alpha }}/{u_{n;1\alpha }},
        \end{array}\]
        where $x_i,y_i(i=1,2,3)$ denote positions of the chosen atoms, $u_{n;i\alpha}$($\alpha=x$ or $y$ for in-plane modes, and $\alpha=z$ for out-of-plane modes) is the component of ${\bf u}_{n}$ on atom $i$ along the $\alpha$ direction. From the above equations, we can get the primitive wave vector as:
        \begin{equation}\label{eq:kxky}
        \left( \begin{array}{l}
        {k_x}\\
        {k_y}
        \end{array} \right) = {\left( {\begin{array}{*{20}{c}}
        {{x_2} - {x_1}}&{{y_2} - {y_1}}\\
        {{x_3} - {x_1}}&{{y_3} - {y_1}}
        \end{array}} \right)^{ - 1}}\left( \begin{array}{l}
         - i\log \left( {{u_{2\alpha }}/{u_{1\alpha }}} \right)\\
         - i\log \left( {{u_{3\alpha }}/{u_{1\alpha }}} \right)
        \end{array} \right).
        \end{equation}
    Therefore, by combining the mode-matching method with Eq. (\ref{eq:kxky}), we can acquire the information of both wave vectors and transmission for an injection mode.\\

 \begin{figure}
    \centering
    \includegraphics[width=0.4\textwidth]{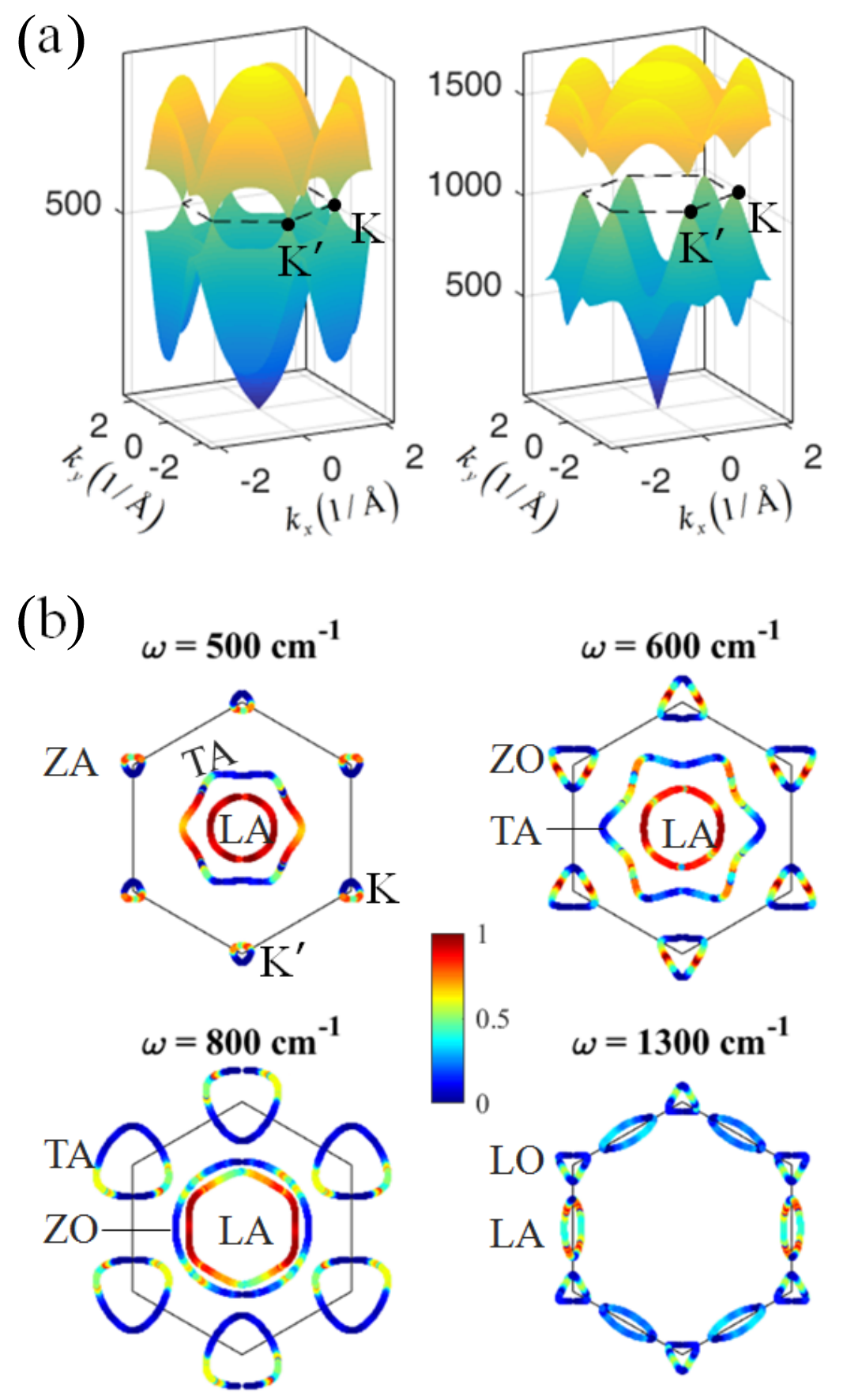}\\
    \caption{
     $\bf k$-resolved transmission function. (a) Schematic plots of $K$-valleys in the ZA/ZO (left panel) and TA/LO (right panel) branches of graphene. The hexagonal dashed lines indicate the first BZ of graphene. (b) $\bf{k}$-resolved transmission at $\omega=500,600,800$ and $1300$ cm$^{-1}$. Transmission of modes that have negative group velocities (left-going) is also drawn by time-reversal symmetry ($\Xi(k_x,k_y)=\Xi(-k_x,-k_y)$) for clarity. Boundaries of the first BZ are marked by black solid lines. }\label{fig2}
    \end{figure}

\section{Valley filtering effect}

For completeness, phonon dispersion of graphene is shown in Fig. \ref{fig1}(b). 
In the left panel of Fig. \ref{fig1}(c), it is demonstrated that BZs are folded in our transport calculations. The unit cell used for the thermal lead L contains 4 primitive cells of graphene, and thus the corresponding BZ is folded. For example, a $\bf k$-vector in the primitive 1$^{\textrm{st}}$ BZ ${\bf k}=(k_x,k_y)=(K_x+N_1{B}_1, K_y+N_2{B}_2)$ ($N_{1,2}=0,1,\cdots$; ${\bf B}_{1,2}$ are reciprocal lattice vectors of the unit cell used for thermal lead L; $|{\bf K}_{x(y)}|<|{\bf B}_{1(2)}|$) is reduced to be ${\bf K}=(K_x,K_y)$ in the folded BZ.

\begin{figure}
    \centering
      \includegraphics[width=0.45\textwidth]{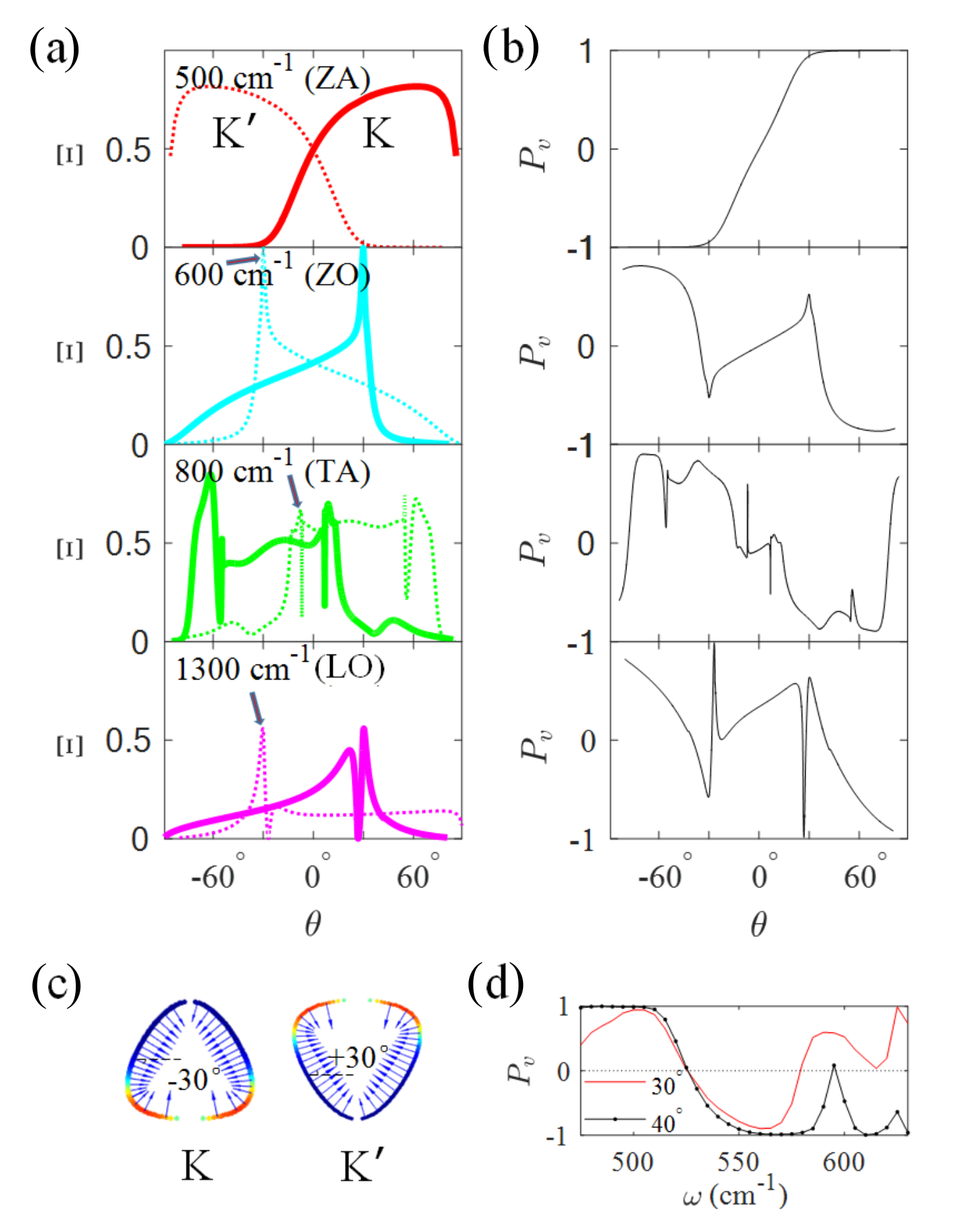}\\
      \caption{Angular dependence of (a) Transmission of valley K (thin solid line) and K' (dotted thick line) phonons and (b) the corresponding valley polarization at 500 (ZA), 600 (ZO), 800 (TA), and 1300 (LO) cm$^{-1}$. (c) Transmission plots with velocity-vectors shown for valley K and K' phonons at 500 cm$^{-1}$. (d) Dependence of valley polarization on phonon wave number of the ZA and ZO branches for an injection angle of 30$^\circ$ (red solid line) and 40$^\circ$ (black dot solid line).
       }\label{fig3}
    \end{figure}

    Figure \ref{fig1}(d,e) demonstrates mode transmission in both the folded BZ and primitive BZ at 200 and 400 cm$^{-1}$.
    In the left panel of Fig. \ref{fig1}(d), the inner two circles in the folded BZ can be attributed to the TA and LA branches directly. Unfolding $(K_x, K_y)$ to primitive $(k_x,k_y)$ in the middle panel, one can observe that the other arcs actually originate from the ZA branch of graphene by comparison with the corresponding contour plot of graphene in the right panel.

    For phonons at 200 cm$^{-1}$, the TA and LA phonons almost fully transmit through the 558 GB; the ZA phonons have the largest transmission at perpendicular injection ($k_y=0$), and smaller transmission at a skew injection angle, which shows a similar trend as previous results in silicon with a high-energy $\Sigma$29 twist GB.\cite{Kimmer_PRB2007} The advantage of unfolding is more clearly seen at 400 cm$^{-1}$. In this case, transmission of the LA and ZA branches is fairly high; by contrast, the TA branch has high transmission when the angle between $k_y$ and $k_x$ is roughly within 30 degree.

    By examining the primitive modes, we are able to explore the valley-selective transmission in the GB system. Illustrated in Fig. \ref{fig2}(a), phonon ``valleys" also exist in the ZA, ZO, TA, and LO branches, which emerge around the Dirac-like points [550 cm$^{-1}$ in Fig. \ref{fig1}(b)] and band extremum (1050 and 1227 cm$^{-1}$) at $K$($K'$). Similarly to the electron case, there are two inequivalent $K$ points, labeled as $K$ and $K'$, at the corners of the $1^{\textrm{st}}$~BZ. 
    Without losing generality, we choose $\omega=$500, 600, 800, 1300 cm$^{-1}$ to illustrate the valley-selective transmission spectrum in Fig.~\ref{fig2}(b).


    For phonons at 500 cm$^{-1}$, TA modes have high transmission when the angle between wave vector $\bf k$ and the transport direction ($x$) is within 30 degree. The ZA valleys around the BZ corners, however, have different features. First, the shape of ZA valleys is triangular instead of circular or hexagonal. Second, transmission of ZA valleys at six corners is classified into two groups: three equivalent ones at $K$ valleys, and the other three equivalent ones at $K'$ valleys. Third, high-transmission modes lie on the horizontal edge of the triangular valleys.

    For phonons at 600 cm$^{-1}$, TA phonons have much lower transmission. The contour shapes of the ZO valleys resemble those of ZA phonon valleys at 500 cm$^{-1}$. However, high transmission modes lie in two different edges in ZO valleys from those edges in ZA valleys at 500 cm$^{-1}$. Transmission of TA valleys at 800 cm$^{-1}$ has similar features to ZA valleys at 500 cm$^{-1}$, and the LO valleys at 1300 cm$^{-1}$ to ZO at 600 cm$^{-1}$.
\begin{figure}
\centering
      \includegraphics[width=0.35\textwidth]{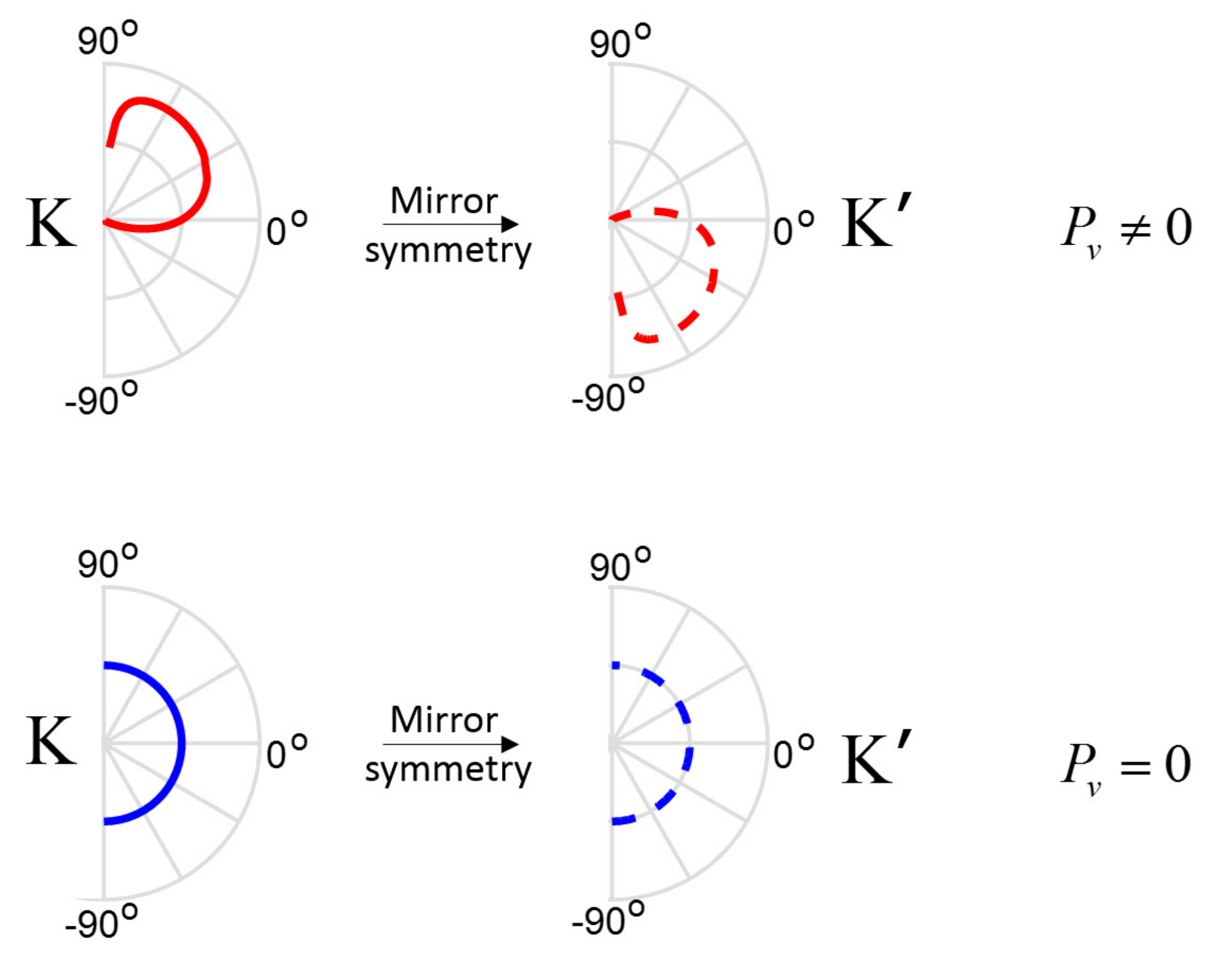}\\
      \caption{Schematic plot of the anisotropy-induced valley-polarized transmission. (Upper panel) Under the mirror symmetry about the $x$ axis, transmission of one phonon valley can be inferred from that of the other valley. Anisotropic transmission naturally leads to valley-polarized transmission. (Lower panel) When transmission of one valley also mirror-symmetric about the $x$ axis, transmission has no valley polarization.
      }\label{fig4new}
    \end{figure}

To get a better idea of the valley filtering effect, we define $\theta=\arctan(v_y/v_x)$ to be the injection angle, which ranges from $-90$ degree to $90$ degree for right-going modes. The angular dependence of valley transmission for those shown in Fig. \ref{fig2}(b) is depicted in Fig. \ref{fig3}(a). Angular dependence of valley transmission shows good symmetry, $\Xi^K(\theta)=\Xi^{K'}(-\theta)$, at $\omega=500$, $600$~cm$^{-1}$, and weak symmetry for $800$ and $1300$ cm$^{-1}$. This deviation may be caused by the slight mirror-symmetry breaking of the optimized structure (the average deviation per atom is about 0.05~\AA). The angles for achieving highest/lowest transmission for two valleys are roughly opposite to each other, $\theta_m(K)\approx-\theta_m(K')$, which further leads to the valley filtering effect of the 558 GB system.

Defining valley polarization of transmitted modes as
 \begin{equation}
 P_v=(\Xi^K-\Xi^{K'})/(\Xi^K+\Xi^{K'}),
  \end{equation}
  we obtain valley polarization at different $\omega$. Figure \ref{fig3}(b) depicts the valley polarization correspondingly to Fig. \ref{fig3}(a). All branches show significant valley polarization. Particularly, the ZA and ZO branches have highest valley polarization ($\pm$100\% and $\pm$86\%) at $|\theta|\to 90^\circ$, which is nearly parallel to the GB. Due to the triangular shape of valley pockets [Fig. \ref{fig2}(b)], transmission of right-going modes mostly occurs among $-90^\circ<\theta<30^\circ$ for one valley and $-30^\circ<\theta<90^\circ$ for the other valley. [Fig. \ref{fig3}(c)] As a consequence, valley polarization for ZA and ZO branches when $|\theta|>30^\circ$ is fairly high because one valley dominates transmission when $|\theta|>30^\circ$. Therefore, the filtering effect is closely related to the anisotropy of phonon valley pockets, implying that such a valley filtering effect is not limited in the 558 GB.

 Also shown in Fig.~\ref{fig3}(c), K and K' valley pockets are nearly inverse symmetric. For a structure which has the mirror symmetry about the $x$ axis (the transport direction),  the angular dependence of valley transmission functions around the K and K' valleys should be $\Xi^K(\theta,\omega)=\Xi^{K'}(-\theta,\omega)$ or $\Xi^K(\delta k_x,\delta k_y,\omega)=\Xi^{K'}(\delta k_x',-\delta k_y',\omega)$, where $\delta {\bf{k}}={\bf{k}}-{\bf{K}}, \delta {\bf{k'}} ={\bf{k'}}-{\bf{K'}}$ are measured from the bottom/top of the valley pockets. Therefore, when the transmission function of one phonon valley lacks the symmetry about $\theta=0$, we obtain $\Xi^K(\theta,\omega)\ne \Xi^{K'}(\theta,\omega)$, which is valley-polarized, as illustrated in Fig.~\ref{fig4new}.

To further find out the filtering energy window for ZA and ZO modes, we plot the valley polarization for $\theta=30^\circ$ and $40^\circ$ in Fig. \ref{fig3}(d). From this plot, two perfect energy windows are found, [475, 510] and [555, 580], within which valley polarization of flexural modes (ZA/ZO) exceeds 95\% when the injection angle reaches $40^\circ$.

\section{Fano-like resonance}
\begin{figure*}
\centering
      \includegraphics[width=0.7\textwidth]{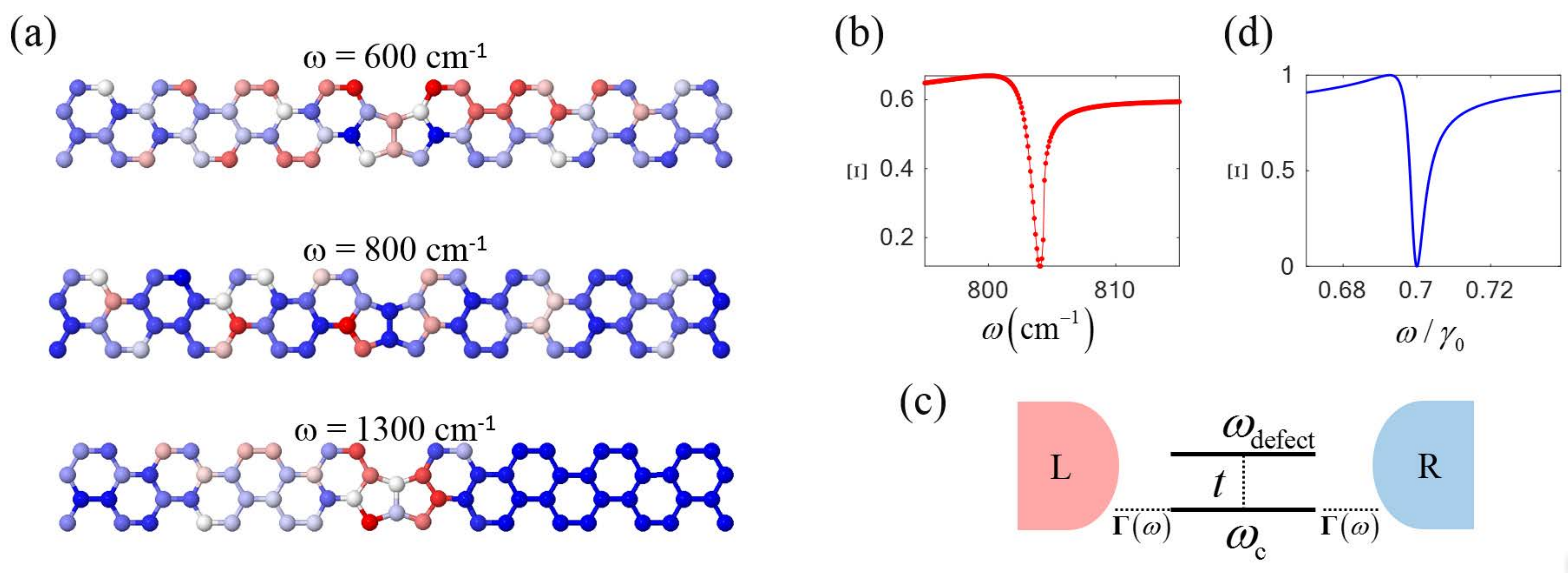}\\
      \caption{ (a) Illustration of the scattering wave functions of most-transmitted modes [indicated by arrows in Fig. \ref{fig3}(a)] at $\omega=600, 800, $ and $1300$ cm$^{-1}$. Effective oscillation amplitudes, which are defined as $u_{\textrm{eff}}=\sqrt{u_x^2+u_y^2+u_z^2}$ are strongest in red and weakest in blue. (b)
      Transmission spectrum around 800 cm$^{-1}$ at $K_y=-0.4008~(2\pi/A_y)$, which leads to the transmission peak indicated by an arrow in Fig. \ref{fig3}(a) at $\omega=800$ cm$^{-1}$. (c) Fano-like resonance model and (d) a numerical result using $\gamma(\omega)=\omega\gamma_0$, $\omega_c=0.84\gamma_0$, $\omega_{\textrm{defect}}=0.7\gamma_0$, $t=0.05\gamma_0^2$.
      }\label{fig4}
    \end{figure*}

    Besides transmission peaks, it is worth noting that there are transmission dips at $\omega=800$ and $1300$~cm$^{-1}$ in Fig.~\ref{fig3}(a). The asymmetric line-shape around the transmission dip and peak implies a Fano-like resonance. To understand the physical origin of the Fano-like resonance, we illustrate the scattering wave functions of highly-transmitted modes [indicated by black arrows in Fig. \ref{fig3}(a)] in  Fig. \ref{fig4}(a). In these plots, the atoms having the highest oscillation amplitude all locate at the GB region. However, the scattering wave function at $\omega=600$ cm$^{-1}$ is less local, which may explain the absence of dips. Based on the observation, we anticipate that the resonance is caused by the interplay between a continuum state from the left thermal lead, $i.e.$, graphene, and a localized state at the grain boundary. Localized quasi-1D metallic states\cite{Batzill_nat2010,ZouXL_small2015,ZouXL_AccChem2015} for electrons at grain boundaries have been found previously. Therefore, it is a reasonable conjecture that local phonon states may also form at the GB.

 The transmission as a function of phonon energy for a fixed $K_y$ mode is further demonstrated in Fig. \ref{fig4}(b). The asymmetry line-shape maintains in the energy domain. Previous theoretical investigations have demonstrated Fano resonances due to the coupling of optical phonon modes,\cite{Lee_PRL2006,Heinz_PRL2012} bound excitons to LO phonons,\cite{ShiSL_JCP2005} and coupling of phonon modes in a 1D nonlinear chain model.\cite{Kim_PRB2001} Inspired by the role of local states in Fano resonance, we build a simple model to interpret the GB system: as shown in Fig. \ref{fig4}(c), there is one local vibration mode $\omega_{\textrm{defect}}$ that interacts with a conducting mode $\omega_c$, which couples directly to two thermal leads.\cite{KimPRL2012} The dynamical matrix of the central region without thermal baths can be written as
    \begin{align}
    {\bm D}_C=\left(
    \begin{array}{cc}
    \omega_c^2 & t \\
    t & \omega_{\textrm{defect}}^2
    \end{array}
    \right),
    \end{align}
    where $t$ represents the interaction of these two modes. The influence of two thermal baths can be counted in using self-energy matrices as
    \begin{equation}
    {\bm\Sigma}^r_{L/R}=-\frac{i}{2} \left(
    \begin{array}{cc}
    \gamma(\omega) & 0\\
    0 & 0
    \end{array}
     \right).
    \end{equation}
  The corresponding bandwidth functions are ${\bm \Gamma}_{L/R}=-2\textrm{Im}({\bm \Sigma}^r_{L/R})$. The retarded Green's function of the central region is
    \begin{equation}
    {\bf G}^r_C=[ (\omega +i\eta)^2 {\bf I}-{\bf D}_C-{\bf \Sigma}_L^r-{\bf \Sigma}_R^r]^{-1}.
    \end{equation}
    And the transmission spectrum can be calculated as\cite{cxb_small2018}
    \begin{align}
    \Xi(\omega) &= \mathrm{Tr} ({\bf \Gamma}_L {\bf G}^r_C {\bf \Gamma}_R {\bf G}^a_C) \\
    &=\frac{{{{\left[ {\gamma \left( \omega  \right)} \right]}^2}}}{{{{[({\omega ^2} - \omega _c^2) - {t^2}/({\omega ^2} - \omega _{{\rm{defect}}}^2)]}^2} + {{\left[ {\gamma \left( \omega  \right)} \right]}^2}}}. \label{eq:trans}
    \end{align}
   Because ${\bf \Sigma}^r(-\omega)={\bf \Sigma}^{r*}(\omega)$,\cite{WangJiansheng_PRE2007,XuYong_PRB2008} $\gamma(\omega)$ should be an odd function of $\omega$. Then, the simplest expression of $\gamma(\omega)$  is $\gamma(\omega)=\omega\gamma_0$, which is actually the wide-band limit model for phonons.
   The expression of $\gamma(\omega)$ is not important, because odd functions such as $\omega^3$, $1/\omega$, $\sin\omega$, and $\omega/(\omega^2+1)$ also lead to similar Fano-like line-shapes.

   Based on Eq.~(\ref{eq:trans}), Fig.~\ref{fig4}(d) shows that when the interaction $t$ is rather weak compared to $\gamma_0^2$, Fano-like resonance manifests around the local mode $\omega_{\textrm{defect}}$. The line-shape shown is partially consistent with the one shown in Fig. \ref{fig4}(b). This simple model demonstrates the existence of resonance and anti-resonance points of transmission as a function of energy. However, it does not explicitly contain angular dependence, which may be further considered by introducing $\bf{k}$-dependence in the model. Our calculations enrich the physics by showing that Fano-like resonance from pure phonon modes also occur in 2D materials. Since dips and peaks emerge at such resonance points, valley polarization can be significantly enhanced around the resonant points.

\section{Conclusion}
In summary, we investigated the phonon transport properties in a 2D graphene sheet which contains an extended zero-angle grain boundary (558 GB). Combining real-space first-principles calculations with the extended mode-matching method to recover the primitive wave vectors of injection modes, we revealed that the 558 GB selectively scatters different phonon modes, leading to valley-selective transport. The valley filtering effect is closely related to the anisotropy of valley pockets of phonons. In particular, the 558 GB can be used as a perfect phonon valley filter for ZA and ZO valleys within [475, 510]  and [555, 580] cm$^{-1}$, respectively. In addition, Fano-like resonance was revealed, which can further enhance the valley polarization of phonons.

Due to crucial computational load, only one type of grain boundary was investigated. However, since the valley filtering effect is closely related to the anisotropy of valley pockets, we anticipate that phonon valley filtering effect can be generally observed in various GB systems and expect that our method will be beneficial in revealing novel properties of phonons in emerging 2D materials.\cite{ZouXL_AccChem2015,XuYong_PRL_2013,NanoLett2013_XZou}\\

\begin{acknowledgments}
The authors want to thank Prof. Xiaolong Zou, Prof. Jian-Sheng Wang, Prof. Xiaohong Zheng, and Dr. Lei Zhang for helpful discussions.
We gratefully acknowledge financial support by NSF-China [Grant Nos. 11704257 (X.C.), 11874035 (Y.X.), and 11704238 (J.W.)], the General Research Fund (Grant No. 17311116), the University Grant Council (Contract No. AoE/P-04/08) of the Government of HKSAR (J.W.), and the NSERC of Canada (H.G.). We thank Calcul Qu\'ebec and Compute Canada for the computation facilities.
\end{acknowledgments}

\appendix
\renewcommand\thefigure{\thesection.\arabic{figure}}
\section{Mode-matching method}
    In this section, we briefly introduce the basic ideas of modified mode-matching method for studying quasi-ballistic transport properties of phonons.

        \subsection{Lippmann-Schwinger Equation}
            Suppose we have an incoming wave function $\psi(\bf r)$ going through the scattering region. To get the wave function across the scattering region, we can use either quantum mechanics methods plus some connection conditions to solve the problem, or the Lippmann-Schwinger (LS) equation. The LS equation states that the scattering wave function $\psi$ is composed of the original incoming wave function ${\psi _0}$ and the scattered wave function as:\cite{Wang_PRB2009}
            \begin{align}
            \psi \left( {\bf{r}} \right)
            & = {\psi _0}\left( {\bf{r}} \right) + \int_{}^{} {G_0^r} \left( {{\bf{r}},{\bf{r}}'} \right)V\left( {{\bf{r}}'} \right)\psi \left( {{\bf{r}}'} \right)d{\bf{r}}'\cr
            & = {\psi _0}\left( {\bf{r}} \right) + \int_{}^{} {{G^r}} \left( {{\bf{r}},{\bf{r}}'} \right)V\left( {{\bf{r}}'} \right){\psi _0}\left( {{\bf{r}}'} \right)d{\bf{r}}',
            \end{align}
            where $V(\bf r' )$ is the scattering potential, $G^r\left( {{\bf{r}},{\bf{r}}'} \right)$ and ${G_0^r}\left( {{\bf{r}},{\bf{r}}'} \right)$ are the propagators of the mode with and without scattering from the central region. Using discrete lattice model, the retarded phonon Green's functions for the system without and with scattering ($G_0^r$ and ${G^r}$) can be defined as
            \begin{align}
            \left[ {{{\left( {\omega  + i\eta } \right)}^2}{\bf I} - {\bf {D_0}}} \right]{\bf G_0}^r & = {\bf I}, \label{G0} \\
            \left[ {{{\left( {\omega  + i\eta } \right)}^2}{\bf I} - {\bf {D_0}} -{\bf  D}'} \right]{{\bf G}^r} &= {\bf I},  \label{G1}
            \end{align}
            where ${\bf D}_0$ is the dynamical matrix of the central part and ${\bf D}'$ is the perturbed dynamical matrix inducing scattering.
            If ${\psi _0}$ is an eigen mode with the energy $E=\hbar \omega$ ($\hbar=1$ hereafter) from the left thermal lead, $\psi$ should also be an eigen state of the two-probe system with the same energy due to the law of energy conservation in quasi-ballistic regime, i.e.,
            \begin{align}
            \left( {{\omega ^2}{\bf I} - {\bf {D_0}}} \right){ \psi _0} = {\bf 0},\label{psi0}\\
            \left( {{\omega ^2}{\bf I} - {\bf {D_0}} - {\bf D}'} \right) \psi  = {\bf 0}.\label{psi1}
            \end{align}
            Combining Eqs.~(\ref{G0}-\ref{psi1}), we have
            \begin{align}
            &\left[ {{{\left( {\omega  + i\eta } \right)}^2}{\bf I} - {\bf {D_0}}} \right]\left( {{\psi _0} + {\bf G}_0^r{\bf D}'\psi } \right)\cr
            =& {\bf D}'\psi = \left( {{\omega ^2} - {{\bf D}_0}} \right)\psi \\
            &\left[ {{{\left( {\omega  + i\eta } \right)}^2}{\bf I} - {\bf {D_0}} -{\bf  D}'} \right]\left( {{\psi _0} + {{\bf G}^r}{\bf D}'{\psi _0}} \right) \cr
             =&  - {\bf D}'{\psi _0} + {\bf D}'{\psi _0} = \bf{0}
            \end{align}
            Therefore, under boundary conditions that
            \begin{align}
            {\left. \psi  \right|_{{\bf D}' = 0}} = {\psi _0}
            \end{align}
            we can express the scattering wave function  $\psi$ in terms of phonon Green's functions as
            \begin{align}
            \psi & = {\psi _0} + {\bf G}_0^r{\bf D}'\psi   \\
                  &= {\psi _0} + {{\bf G}^r}{\bf D}'{\psi _0}. \label{eq:psi=psi0+GDpsi0}
            \end{align}

        \subsection{Transmission}
            With the information of scattering states, it is straight forward to calculate the transmission function. Combining Eq.~(\ref{eq:psi=psi0+GDpsi0}) and the Dyson Equation,
            \begin{align}
            {\bf G}^r = {{\bf G}_0^r}
             + {\bf G}^r{\bf D}'{{\bf G}_0^r},
            \end{align}
            we can reformulate $\psi$ in terms of phonon Green's functions as
            \begin{equation}
            \psi  = {\bf G}^r{\left( {{{\bf G}_0^r}} \right)^{ - 1}}{\psi _0}. \label{eq:psi=G*G0niPsi0}
            \end{equation}
            Further by decomposing $\psi$ into eigen channels of the right thermal lead, we can obtain mode-resolved transmission. In detail, the right-going modes from the left thermal lead are denoted as
            \begin{align}
            {{\bf{u}}_{Ln}}( + ),\;n = 1,...,{\rm{M}},
            \end{align}
            where $M$ represents the degree of freedom in a principal layer. Going through the central scattering region, ${{\bf{u}}_{Ln}}( + )$ becomes
            \begin{align}
            {\psi _{l}} = {\bf G}_{n,0}^r{\left( {{\bf G}^{(0)r}_{L;0,0}} \right)^{ - 1}}{{\bf{u}}_{Ln}}( + )
            \end{align}
            at the $l^{\textrm{th}}$ principal layer. Here, the $0^{\textrm{th}}$ layer is the surface principal layer of the left thermal lead and
            ${ {\bf G}^{(0),r}_{L;0,0}}$ is the retarded Green's function at the $0^{\textrm{th}}$ principal layer of the semi-infinite left thermal lead, which hosts eigenmodes $\{ {\bf{u}}_{Ln}( + )\}_{n=1}^M$.
            Decomposing the scattering wave function at the surface layer of the right thermal lead, $\psi_{N+1}$, into right-going eigenmodes of the right thermal lead, we have
            \begin{align}
            {\psi _{N+ 1}} = \sum\limits_m^{} {{\tau _{nm}}{{\bf{u}}_{Rm}}\left(  +  \right)}.
            \end{align}
            Then, we get elements of the transmission amplitude matrix as
            \begin{align}
            {t_{mn}}
            & = \sqrt {\frac{{{v_m}{a_L}}}{{{v_n}{a_R}}}} {\tau _{mn}}\cr
            &= \sqrt {\frac{{{v_m}{a_L}}}{{{v_n}{a_R}}}} \tilde{\bf{ u}}_{Rm}^\dag \left(  +  \right){\bf G}_{N + 1,0}^r{\left( {{\bf G}_{L;0,0}^{(0)r}} \right)^{ - 1}}{{\bf{u}}_{Ln}}( + ), \cr \label{eq:tmn}
            \end{align}
            where $v_{m(n)}$ is the group velocity of mode $m(n)$, $a_{L/R}$ is the principal-layer lattice constant of the $L/R$ thermal leads, and dual functions $\tilde{\bf{ u}}_{Rm} \left(  +  \right)$ satisfy
            \begin{align}
            \tilde{\bf{ u}}_{Rm}^\dag \left(  +  \right){\bf{u}}_{Rn}^{}\left(  +  \right) = {\delta _{mn}}.
            \end{align}
            From Eq.~(\ref{eq:tmn}), it is clear that transmission coefficients of phonon modes going from the left to the right thermal lead can be obtained if we have the information of eigen states of both thermal leads, phonon Green's function of the central scattering region, and surface phonon Green's function of the left thermal lead.

\subsection{Obtaining incoming modes at a given energy}
            In this section, we shall discuss how to get eigen phonon modes at a given energy for a periodic system. Eigen modes of the system can be solved according to the following equation as:
            \begin{align}
            {\bf D}{\psi _0}
            &= \left( {\begin{array}{*{20}{c}}
             \ddots & \ddots & \ddots & \ddots & \ddots &{}&{}\\
            {}&{\bf{0}}&{{{\bf{D}}_{ - 1}}}&{{{\bf{D}}_0}}&{{{\bf{D}}_1}}&{\bf{0}}&{}\\
            {}&{}& \ddots & \ddots & \ddots & \ddots & \ddots
            \end{array}} \right){\psi _0}\cr
            &= {\omega_0 ^2}{\psi _0}, \label{eq:Dpsi=w2psi}
            \end{align}
            where ${\bf D}$ is the dynamical matrix of the system. This equation leads to a complete set of orthogonal eigenstates with the eigenvalue $\omega_0^2$. We may write the wave function in real space in terms of local wave functions at each principal layers (or unit cells) as
            \[{\psi _0} = \left( \begin{array}{l}
             \vdots \\
            {{\bf {c}}_{ - 1}}\\
            {{\bf {c}}_0}\\
            {{\bf {c}}_1}\\
             \vdots
            \end{array} \right).
            \]
            Substituting the above equation into Eq.~(\ref{eq:Dpsi=w2psi}), we have a chain-like relationship between adjacent principal layers\cite{Ando_PRB_1991}:
            \begin{align}
            {{\bf D}_{ - 1}}{{\bf {c}}_{i - 1}} + {{\bf D}_0}{{\bf {c}}_i} + {{\bf D}_1}{{\bf {c}}_{i + 1}} = {\omega_0 ^2}{{\bf {c}}_i}. \label{eq:chain}
            \end{align}
            Because this is a periodic system, the Bloch's theorem guarantees that wave functions at adjacent principal layers differ by only a phase factor:
            \begin{align}
            {{\bf {c}}_{i \pm 1}} = \lambda_n^{\pm 1}{{\bf {c}}_i}, \label{eq:bloch}
            \end{align}
            with $\lambda_n={e^{  iK_na}}$. ($a$ is the lattice constant, which is $A_1$ in our calculation.) Replacing ${\bf {c}}_{i\pm1}$ in Eq.~(\ref{eq:chain}) by ${\bf {c}}_i$ according to Eq.~(\ref{eq:bloch}) and using an eigenmode ${\bf u}_n$ instead, we get
            \begin{align}
            \left( {{{\bf D}_{ - 1}}{\lambda_n ^{ - 1}} + {{\bf D}_0} + {{\bf D}_1}\lambda_n } \right){{\bf {u}}_n} = {\omega_0 ^2}{{\bf {u}}_n}, \label{eq:chain2}
            \end{align}
            which is a quadratic equation for $\lambda_n$. To facilitate the solution, we can rewrite the above equation into a linear one:
            \[
            \left( {{{\bf D}_{ - 1}}{{\bf {u}}_n} + {{\bf D}_0}\lambda_n {{\bf {u}}_n}} \right) = \lambda_n \left( {{\omega_0 ^2}{{\bf {u}}_n} - {{\bf D}_1}\lambda_n {{\bf {u}}_n}} \right)
            \]
            or in the form of matrix:
            \begin{align}\label{eq:modeSolve}
            &\left( {\begin{array}{*{20}{c}}
            {{{\bf D}_{ - 1}}}&{{{\bf D}_0}}\\
            {\bf 0}& {\bf I}_{N\times N}
            \end{array}} \right)\left( \begin{array}{c}
            {{\bf {u}}_n}\\
            \lambda_n {{\bf {u}}_n}
            \end{array} \right) \cr
            =& \lambda_n \left( {\begin{array}{*{20}{c}}
            {{\omega_0 ^2}{\bf I}_{N\times N}}&{ - {{\bf D}_1}}\\
            {\bf I}_{N\times N}&{\bf 0}
            \end{array}} \right)\left( \begin{array}{c}
            {{\bf {u}}_n}\\
            \lambda_n {{\bf {u}}_n}
            \end{array} \right).
            \end{align}
            Note that the matrix form is not unique. Then, we can solve the equivalent equation
            \begin{align}\label{eq:Apsi=lambda B psi}
            {\bf A\Psi}  = \lambda_n {\bf B\Psi} ,
            \end{align}
            where
            \begin{align}
            \bf A &=  \left( {\begin{array}{*{20}{c}}
            {{{\bf D}_{ - 1}}}&{{{\bf D}_0}}\\
            {\bf 0} & {\bf I}
            \end{array}} \right),\\
            \bf  B &= \bf \left( {\begin{array}{*{20}{c}}
            {{\omega_0 ^2}{\bf I}}&{ - {{\bf D}_1}}\\
            {\bf I}&{\bf 0}
            \end{array}} \right),\\
            \bf \Psi & = \left( \begin{array}{c}
            {{\bf {u}}_n}\\
            \lambda_n {{\bf {u}}_n}
            \end{array} \right),
            \end{align}
            and get 2$M$ eigenmodes and the corresponding eigenvalues. Among the 2$M$ eigenmodes, there are $M$ right-going and $M$ left-going modes. And both of them may contain propagating and evanescent modes, judging by the value of $\lambda_n$:\cite{Kelly_prb_2005} (1) if $|\lambda_n|<1$ or $>1$, ${{\bf {u}}_n}$ is an evanescent mode; (2) if $|\lambda_n|=1$, ${{\bf {u}}_n}$ is a Bloch propagating mode.

     \subsection{Velocity of incoming modes}
        Differentiating Eq.~(\ref{eq:chain2}) over $K_n$, we have
        \begin{align}
        \left( {ia {\bf D}_1^{}{e^{iK_na}} - ia{\bf D}_{ - 1}^{}{e^{ - iK_na}}} \right){{\bf {u}}_n} = \frac{{2\omega_0 {\textrm{d}}\omega_0 }}{{{\textrm{d}}K_n}}{{\bf {u}}_n}.
        \end{align}
        Therefore, the group velocity of a normalized propagating eigenmode is ($\hbar=1$)
        \begin{align}
        {v_n} & = \frac{{{\textrm{d}}\omega_0 }}{{{\textrm{d}}K_n}} = \frac{1}{{2\omega_0 }}{\bf {u}}_n^\dag \left( {ia{\bf D}_1^{}{e^{iK_na}} - ia{\bf D}_{ - 1}^{}{e^{ - iK_na}}} \right){{\bf {u}}_n}\cr
         & =  - \frac{a}{\omega_0 }\mathrm{Im}\left( {{\bf {u}}_n^\dag {\bf D}_1^{}{e^{iK_na}}{{\bf {u}}_n}} \right) \cr
         & =  - \frac{a}{\omega _0}{\mathrm {Im}}\left( {\lambda_n {\bf {u}}_n^\dag {\bf D}_1^{}{{\bf {u}}_n}} \right).
        \end{align}
        There is an extra factor of $1/\omega_0$ compared to the electron case.\cite{Kelly_prb_2005,Savito_PRB_2008} Considering that $\omega_0$ is a given constant when obtaining mode-resolved transmission spectrum in Eq.~(\ref{eq:tmn}), we may use
        \begin{align}
        {{\tilde v}_n} =  - a{\mathrm{Im}}\left( {{\lambda _n}{\bf {u}}_n^\dag {\bf D}_1^{}{{\bf {u}}_n}} \right)
        \end{align}
        instead of $v_n$ in Eq.~(\ref{eq:tmn}) for calculating transmission.

 \subsection{Effective Dynamical Matrix}
    To get mode-resolved transmission, we constructed the effective dynamical matrix at each $K_y$ point as:\cite{ZhangGang_PRB_2015,Fisher_PRB2017}
      \begin{equation}
    {{\bf D}_{n_x}}\left( {{K_y}} \right) = \sum\limits_{n_y=0,\pm 1}^{} {{\bf D}\left( {n_x,n_y} \right){e^{in_y{K_y}{A_y}}}},
    \end{equation}
    where $A_y$ is the lattice constant of the transport system along the $y$-direction, and $n_x$($n_y$) marks the position of (adjacent) unit cells. In our model, $A_y=A_2\approx4.88$~\AA~. Using the effective 1D dynamical matrix, we obtained $K_x$-resolved transmission at each $K_y$ point based on the mode-matching method.\cite{WangJian_JAP_2009,ZhangGang_PRB_2015,cxb_prb_2015}


%

\end{document}